\documentclass[%
superscriptaddress,
preprint,
 amsmath,amssymb,
 aps,
pre,
floatfix,
]{revtex4-2}

\usepackage{graphicx}
\usepackage{dcolumn}

\usepackage{bm}
\usepackage{hyperref}

\begin{document}

\title{Chaotic variability in a model of coupled ice streams}
\date{\today{}}
\author{Kolja Kypke}
\affiliation{Department of Mathematics and Statistics, University of Guelph, Guelph, Canada}
\affiliation{Physics of Ice, Climate and Earth, Niels Bohr Institute, University of Copenhagen}
\author{Peter Ashwin}
\affiliation{Department of Mathematics and Statistics, University of Exeter}
\author{Peter Ditlevsen}
\affiliation{Physics of Ice, Climate and Earth, Niels Bohr Institute, University of Copenhagen}

\begin{abstract}
    Regions of fast-flowing ice in ice sheets, known as ice streams, have been theorized to be able to exhibit build-up/surge oscillatory variability due to thermomechanical coupling at the base of the ice.
    A simple model of three coupled ice streams is constructed to replicate the spatial configuration of a single ice stream being bisected into two termini.
    The model is constructed to mimic existing branching ice streams in northern Greenland.
    This model is shown to exhibit both steady-flow and build-up/surge oscillations.
    Further, the variability can be chaotic due to the nonlinear coupling of three incommensurate frequencies.
    This provides a mode of chaotic internal variability for ice sheets that contain these types of ice streams.
\end{abstract}

\maketitle

\section{Introduction}

Fast variations in ice sheet volumes are most commonly assumed to be driven by external components of the climate forcing the ice sheet.
This is because the dynamics of ice sheets are dominated by the slow deformational creep of an extremely viscous fluid.
However, ice sheets possess a mode of fast flow driven by the thermomechanical coupling of the ice sheet to its base.
Due to the immense overburden pressure of the ice, the base can find itself at the pressure melting point such that the base of the ice is liquid water.
The water under the ice sheet promotes rapid movement of the ice, generally in topographically confined channels, known as ice streams.
Important to this process are the characteristics of the bedrock under the ice.
The water can either lubricate the ice sheet base causing it to slide on top of hard bedrock (basal sliding), or by saturating a weak till (unconsolidated sediment of rocks, mud and ice or water), causing it to shear close to the bedrock \cite{bennett_ice_2003}.
Ice-sediment interactions in marine-terminating glaciers can also play a factor \cite{brinkerhoff_sediment_2017}.

Ice streams have been shown to transport large quantities of ice over relatively short periods of time in ice sheet model simulations of the Laurentide ice sheet (LIS) that covered parts of North America during the Last Glacial Period (LGP)  \cite{calov_largescale_2002, papa_intermittent_2006, calov_results_2010, roberts_role_2016, schannwell_sensitivity_2023}.
These ice streams are theorized to be the cause of Heinrich events (HE), which correspond to large amounts of ice-rafted debris found in the marine sediment records of the LGP \cite{heinrich_origin_1988, broecker_origin_1992}.
These HE are marked for their abrupt nature and are sometimes synchronous with the Dansgaard-Oeschger events seen in the ice-core record of Greenland \cite{dansgaard_evidence_1993}.
Whether they are externally forced or occur spontaneously due to the internal variability of the ice sheet is still under contention \cite{alvarez-solas_iceberg_2013}.

Some of the aforementioned model studies of the LIS display oscillations in ice sheet volume corresponding to HEs under constant external forcing.
These oscillations are irregular in period, also described as quasiperiodic \cite{payne_limit_1995}, as expected due to the size and complexity of the ice sheet.
To study these oscillations, Robel et al.~\cite{robel_dynamics_2013} (hereafter R13) created a one-dimensional conceptual model of an ice stream.
This model experiences a subcritical Hopf bifurcation between steady-streaming and oscillations of the ice stream thickness and results in bistability and hysteresis for a range of ice surface temperatures.
However, this model is unable to replicate the quasiperiodic or even chaotic variability observed in the comprehensive models.
To address this, the model has been expanded with a stochastic component \cite{mantelli_stochastic_2016}.
In this case, the noise represents fast, small-scale processes such as atmospheric variability in surface temperature and accumulation rates.
Realizations of the resulting model randomly alternate between steady streaming and oscillations due to noise-induced tipping from one mode to the other.
Still, this does not address the irregularity observed in  the comprehensive models subjected to a constant external forcing.

Recently, a model study of the present-day Greenland ice sheet has displayed ice-stream oscillations similar in period to those of R13 under a constant external temperature forcing \cite{kypke_chaotic_2025}.
Similar to the LIS simulations, the period and amplitude of the observed oscillations are highly irregular despite the external forcing containing no stochastic or chaotic component.
The simulations also experienced a tipping from an ice-covered to an ice-free Greenland, albeit in an unpredictable manner.
The ice-stream oscillations are evidenced to delay the tipping of the ice sheet and are theorized to be chaotic transients \cite{lai_transient_2011}.
The resulting lack of predictability of both the tipping point and tipping time in that study is due to the chaotic internal dynamics themselves, rather than an external stochastic forcing.

In the present study, a conceptual model of ice streams displaying chaotic variability is constructed.
Building upon the model of R13, it is extended to include nonlinear coupling of nearby ice streams.
This coupling is inspired by the configuration of the oscillating ice stream seen in Kypke et al.~\cite{kypke_chaotic_2025}.
What are originally two individual ice streams that behave independently retreat inland under an external temperature perturbation such that their spatial proximity allows them to influence each other, effectively coupling them.
Using this model, it is demonstrated how certain parameters affect the overall variability, and how chaotic modes and chaotic transients can arise solely from ice stream dynamics.
While the setup of this conceptual model does not exactly match the conditions and configuration of the comprehensive ice sheet model simulations on which it is based, there is a suitable similarity to be able to explore the situations that can lead the system to exhibit chaos.

\section{The R13 ice stream model}

\begin{figure}
\includegraphics[width=0.9\linewidth]{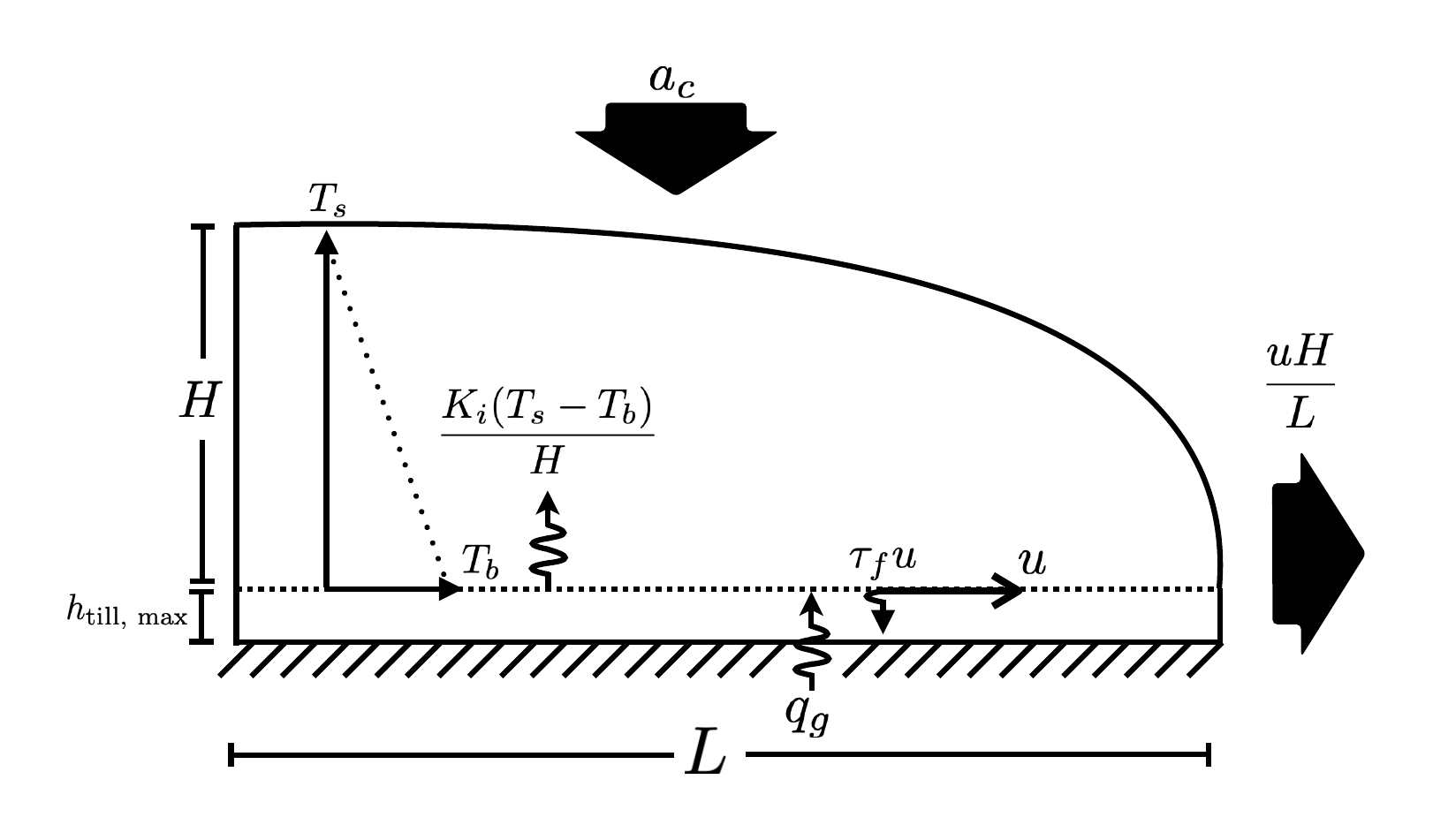}
\includegraphics[width=0.9\linewidth]{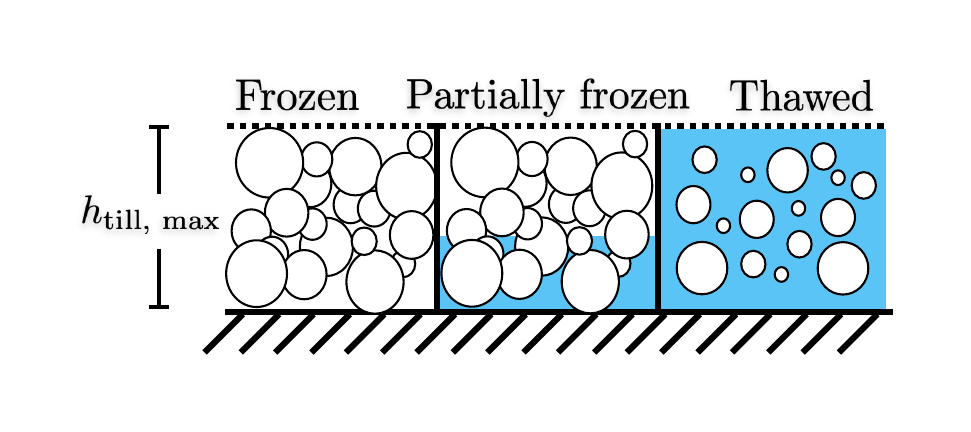}
\caption{(Top) Schematic of the fluxes for the R13 model of a single ice stream. Broad arrows represent mass fluxes and wiggly arrows represent energy fluxes. (Bottom) Schematic of the three cases of the till.}
\label{fig:fig01}
\end{figure}

The R13  model of ice stream temporal variability assumes an idealized geometry of an ice stream with constant width and length, thus only varying height.
The variables of the model describe average values and are reduced to a single spatial dimension of ice thickness.
The upper panel of Fig. \ref{fig:fig01} shows a cross section of the ice stream, displaying the mass and energy fluxes.
The four prognostic variables of the model are ice stream ice thickness $H$, ice sheet basal temperature $T_b$, the thickness of the solids in the unfrozen section of till $h_{\text{till},s}$, and till void ratio $e$, defined below. 
The void ratio is the fraction $e = \frac{V_w}{V_s}$, where $V_w$ is the volume of water and $V_s$ is the volume of solids in the till such that the total till volume is $V_T = V_w + V_s$. 
It can take on any value greater than zero and is related to the porosity $\phi \in [0,1)$ by the relationship 
\begin{equation}
    \phi = \frac{e}{1+e}.
\end{equation}
The thickness of the water in the till is given as $h_{\text{till}, w} =\phi h_{\text{till, max}}$ and is related to the void ratio and thickness of the solids in the unfrozen till by $h_{\text{till}, w} = e h_{\text{till}, s}$.
The basal velocity $u$ is a diagnostic variable determined by the driving stress $\tau_d$ and the basal frictional stress $\tau_f$, which depends on the till water content.
The model has three cases for the state of the basal till, seen in the lower panel of Fig. \ref{fig:fig01} and described below.

The change in ice thickness is
\begin{equation}
\frac{d H}{dt} = a_c - \frac{Hu_b}{L},
\label{eqn:dhdt}
\end{equation}
that is, a balance between the mass gained via accumulation due to precipitation $a_c$ and the mass lost due to sliding, assuming ice thickness vanishes at the end of the ice stream.
The governing equations for $e$, $h_{\text{till},s}$, and $T_b$ depend on the state of the till, being either frozen (A), partially frozen (B), or thawed (C).

\subsection{Frozen till}
The first state is completely frozen till: the unfrozen till height $h_{\text{till},s} = 0$ and either $T_b \geq 0$ (for ease of calculation, $T_b$ is positive for negative temperatures) or it is 0 and the basal melt rate is negative.
In this case, there is no basal sliding,
\begin{equation}
u = 0.
\end{equation}
The void ratio does not change, i.e. no water in the till is melting or freezing,
\begin{equation}
\frac{d e}{dt} = 0,
\end{equation}
such that the unfrozen till thickness stays at 0,
\begin{equation}
\frac{d h_{\text{till},s}}{dt} = 0.
\end{equation}
All heat fluxes go to increasing or decreasing the temperature gradient at the base of the ice,
\begin{equation}
\frac{d T_b}{dt} = -\frac{1}{C_{i} h_b }\bigg(\tau_f u + q_g - \frac{K_i (T_s - T_b)}{H} \bigg).
\end{equation}
The term in brackets, the melt rate, depends on the difference between the geothermal heat flux $q_g$ and the energy diffusing to the surface of the ice as determined by the gradient between the surface temperature $T_s$ and basal temperature, as well as any heat generated by friction due to sliding.
The prefactor $C_{i}$ is the heat capacity of ice and $h_b$ is the thickness of the basal ice layer used to calculate the temperature gradient.

\subsection{Partially frozen till}
The second case is partially frozen till. 
In this case, the basal temperature is $0 ^{\circ}$C, and the void ratio is at a minimum value, $ e = e_c$, known as the till consolidation void ratio \cite{tulaczyk_basal_2000}.
The unfrozen till thickness is greater than zero but must be either less than its maximum value, or it is at its maximum value and with a negative melt rate. 
In this case, there is again no basal sliding since it is still partially frozen,
\begin{equation}
u = 0,
\end{equation}
the void ratio does not change,
\begin{equation}
\frac{d e}{dt} = 0.
\end{equation}
The unfrozen till thickness changes depending on the melt rate,
\begin{equation}
\frac{d h_{\text{till},s}}{dt}= \frac{1}{L_f \rho_i}\bigg(\tau_f u + q_g - \frac{K_i (T_s - T_b)}{H} \bigg),
\end{equation}
where $L_f$ is the latent heat of fusion of ice and $\rho_i$ is the density of ice.
As all the energy is being used to melt or refreeze the till, the basal temperature stays at 0$^{\circ}$C
\begin{equation}
\frac{d T_b}{dt} = 0.
\end{equation}

\subsection{Thawed till}
The final case is where the till is thawed. The unfrozen till thickness will be maximum, and any energy is used to increase or decrease the void ratio. 
The basal velocity in nonzero and is given as
\begin{equation}
u = \frac{A_f W^{n+1}}{4^n(n+1)H^n}\max{[\tau_d-\tau_f, 0]}^n,
\end{equation}
where the driving stress $\tau_d$ depends on the ice thickness and surface slope and the basal stress $\tau_f$ depends on the water content.
It is equal to $\infty$ for fully or partially frozen till (resulting in $u=0$ in cases A and B) and decreases to zero with increasing till water content as $\tau_f = \tau_0\exp{[-c e]}$. 
The parameter $A_g$ is the rate factor and $n=3$ the exponent of Glen's Flow law \cite{paterson_structure_1994}.
The void ratio changes depending on the melt rate,
\begin{equation}
\frac{d e}{dt} = \frac{1}{h_{\text{till},s}L_f\rho_i}\bigg(\tau_f u + q_g - \frac{K_i (T_s - T_b)}{H} \bigg),
\end{equation}
the unfrozen till thickness does not change and stays maximal,
\begin{equation}
\frac{d h_{\text{till},s}}{dt} = 0,
\end{equation}
 and the basal temperature stays at 0$^{\circ}$C,
\begin{equation}
\frac{d T_b}{dt} = 0.
\end{equation}
A large, increasing void ratio for a fixed total till thickness represents the solids gradually being replaced until the entire till is just a layer of water.
Physically, a maximum void ratio that depends on the till properties would be expected, and any water beyond this is somehow drained from the system.
A large void ratio is not an issue in this simplified model however, as the only component that depends on the void ratio is the basal friction $\tau_f$ which saturates to near zero very quickly as the void ratio increases.

\section{Split ice stream model}
\begin{figure}
\includegraphics[width=\linewidth]{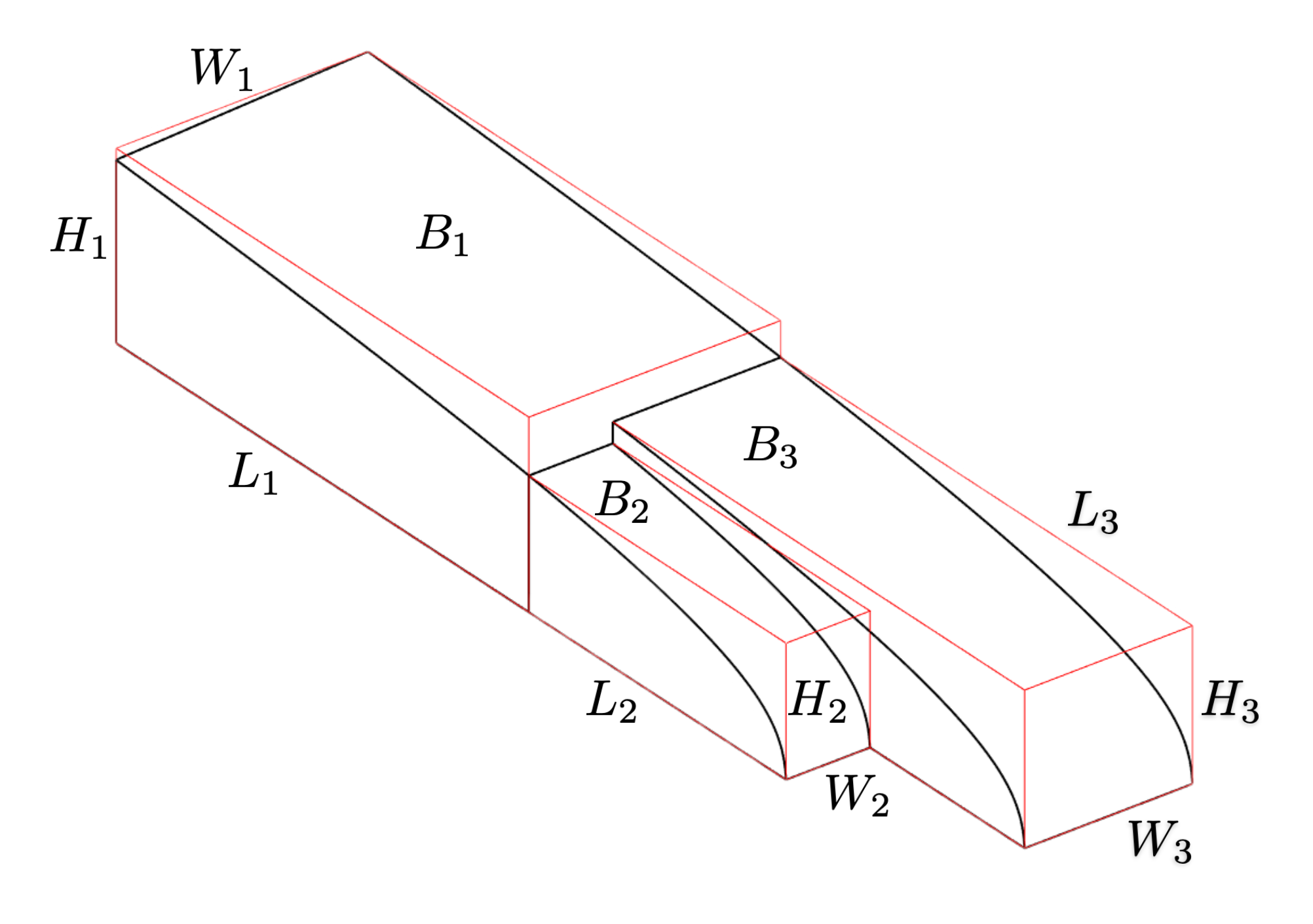}
\caption{Schematic of the geometry considered in the three-box split ice stream model. B1 gains volume from accumulation and loses volume to B2 and B3 due to streaming flow. B2 and B3 gain volume from accumulation and as volume flux from B1, and lose volume due to streaming flow.}
\label{fig:fig02}
\end{figure}

The R13 model is modified to represent a configuration of ice streams applied in Kypke et al. \cite{kypke_chaotic_2025}.
There a single large ice stream is split into two termini.
These ice stream sections have differing basal properties and geometries, the subsequent oscillations will have different amplitudes and periods.
This is modelled as a coupling between three ice streams, each described by the R13 model.
The coupling is achieved by designating three box models and coupling them by conserving volume between one upstream box (B1) and two smaller downstream boxes (B2 and B3) as in Fig. \ref{fig:fig02}.
The figure is not to scale: the lengths are on the order of hundreds of kilometres and the widths on the order of tens of kilometres, while the height is on the order of a kilometre.

For the upstream box, the change in ice thickness due to basal sliding depends on the difference between the height of B1 and the heights of B2 and B3.
This difference is a linear combination of the heights of B2 and B3, weighted by their relative widths.
\begin{equation}
  \frac{dH_1}{dt}  = a_c  -\frac{(H_1 - \frac{W_2}{W_1}H_2 -\frac{W_3}{W_1}H_3 )u_{b,1}}{L_1}
\end{equation}
The widths of B2 and B3 sum up to the width of B1 ($W_1 = W_2+W_3$), but otherwise the rest of the parameters are unconstrained.
Coupling between the boxes also occurs via modulation of the driving stress.
Since the driving stress depends on the surface slope, the driving stress of the upstream B1 is modified to depend on the difference between its height and the height of the downstream boxes in the same manner as described above,
\begin{equation}
\tau_{d1} = \rho_i g \frac{H_1}{L_1} \bigg( H_1 - \frac{W_2}{W_1}H_2 -\frac{W_3}{W_1}H_3 \bigg).
\end{equation}
B2 and B3 are downstream of B1, so any volume loss due to basal sliding of the latter enters the former, each getting a fraction determined by their width.
To keep the coupling unidirectional, the condition is set that the volume flux is only nonzero if the change in volume of B1 is negative.
This prevents the case where B2 and B3 lose volume while B1 is gaining volume.
Explicitly, this means 
\begin{equation}
L_iW_i\frac{dH_i}{dt} =\begin{cases}
 L_iW_ia_c -W_iH_iu_{b,i} - \frac{W_i}{W_1}\frac{dV_1}{dt} & \text{if} \ \ \ \ \frac{dV_1}{dt} <0\\
  L_iW_ia_c -W_iH_iu_{b,i} & \text{otherwise}
\end{cases}
\end{equation}
for $i= 2,3$.

The basal velocity scales most strongly with three parameters: the rate factor $A_g$, the box width, and the box length through the driving stress.
The rate factor itself depends on the temperature of the ice, and can vary by multiple orders of magnitude, from $6.8 \times 10^{-24}$ Pa$^{-3}$ s$^{-1}$ at 0$^{\circ}$C to $3.6 \times 10^{-27}$ Pa$^{-3}$ s$^{-1}$ at -50$^{\circ}$C \cite{paterson_structure_1994}.
The larger the rate factor, the softer the ice, and the faster it can flow.
Since the temperature of the entire ice column is not specified in the model, a constant value of $5 \times 10^{-16}$ as in R13 is used, which corresponds to a temperature of around -10$^{\circ}$C.
The basal velocity scales with the width to the fourth power, but otherwise, the only other appearance of this parameter in the model is in determining the relative strength of the coupling of the driving stress of B1 to B2 and B3.
The primary effect of the length of the box on the basal velocity is to determine the flow regime.
The equilibrium ice thickness in the steady flow regime, derived from equation (\ref{eqn:dhdt}), depends on the length and the basal velocity such that it ultimately scales linearly with length. 
A larger $L$ means a larger ice thickness, which increases the basal melt rate and thus tends towards the steady-flow regime.

As in R13, the geothermal heat flux and surface temperature also determine the flow regime in opposite ways.
A large geothermal heat flux increases the basal melt rate, whereas a more negative surface temperature increases the conduction of heat from the base upwards through the ice sheet, decreasing the basal melt rate.
Depending on the size of the ice stream boxes we may expect a smaller geothermal heat flux in the upstream box, which is closer to the centre of the ice sheet. 
However, a constant value across the boxes is assumed.
The surface temperature is also assumed to be more negative in the upstream box, primarily to represent a lower temperature due to the higher altitude as a result of a larger ice sheet thickness.
Similarly, the accumulation rate in the upstream box is larger to represent less surface melt at lower temperatures.

\section{Chaotic Variability and Routes to Chaos}

\begin{figure}
    	\includegraphics[width=\linewidth]{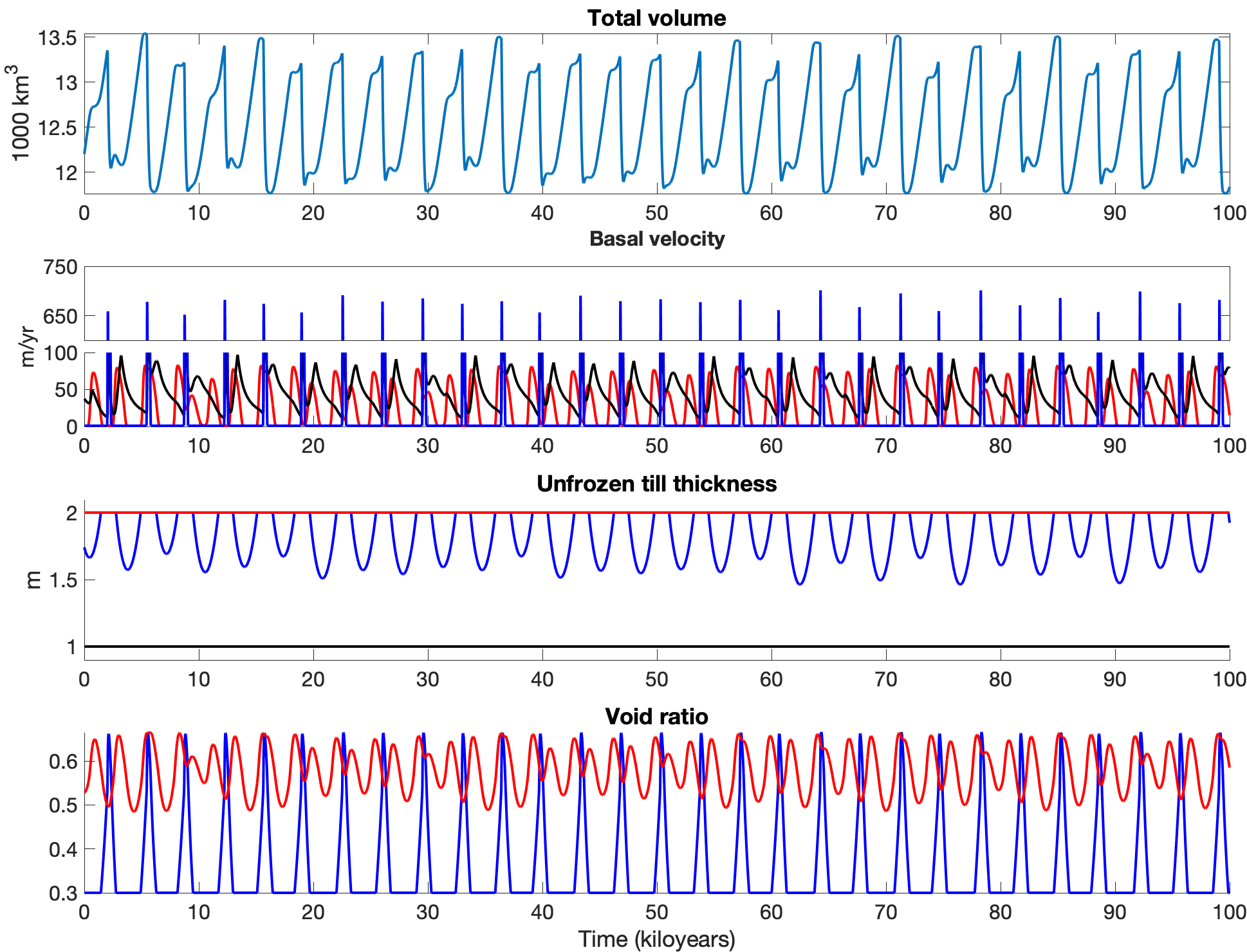}
\caption{Time series of variables for the individual boxes in the chaotic regime. Black curves are B1 values, blue lines are B2, and red lines are B3. Parameter values are given in Tables 1 and 2 in the Appendix, with $T_{s,2} =15.085^{\circ}$C. The void ratio of B1 is arbitrarily large, so it is omitted for readability.}
\label{fig:fig03}
\end{figure}

The split ice stream model displays all the same modes of variability as in the single ice stream model R13: steady flow as well as weak and strong build-up/surge oscillations.
In addition, it is able to exhibit chaotic variability, an example of which is presented in this section.
Figure \ref{fig:fig03} shows a simulation for a set of parameters where the three boxes are in the three different modes:
B1 is in the steady-flow mode, characterized by the arbitrarily large void ratio (not pictured) and small but nonzero basal velocity. 
Spikes in this basal velocity occur due to the coupling to B2, where strong surging occurs.
This build-up/surging mode in B2 is due to occasional refreezing of the till, identified by a basal velocity that becomes zero when the void ratio becomes minimal and the unfrozen till thickness becomes less than maximal.
B3, on the other hand, is in the regime of weaker build-up/surge oscillations.
The unfrozen till thickness is maximal, but the void ratio varies along with the basal velocity.

\begin{figure}[ht]
\includegraphics[width=0.8\linewidth]{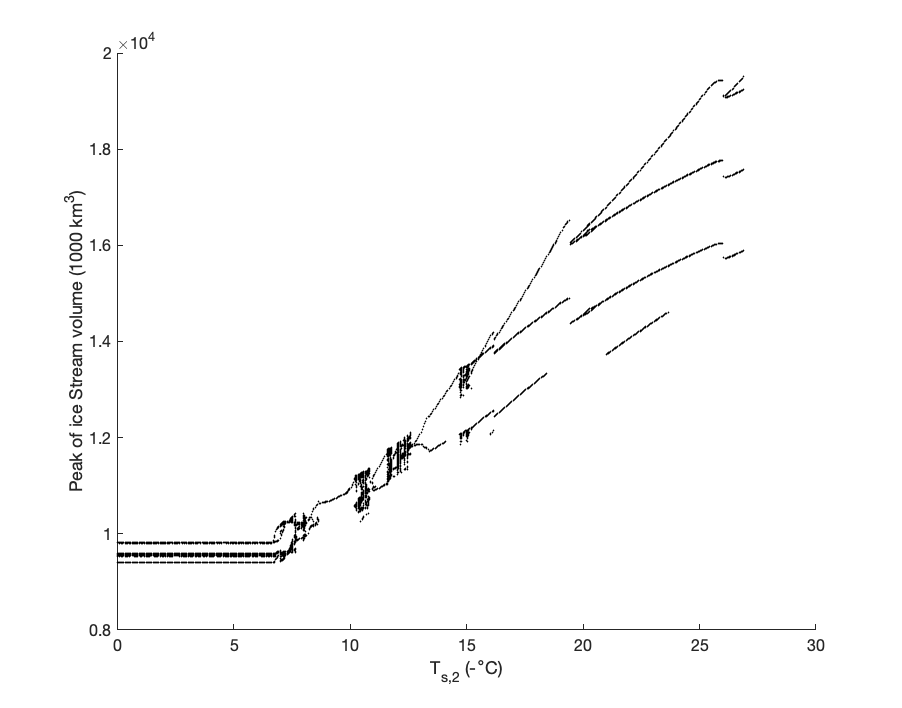}
\quad
\includegraphics[width=0.8\linewidth]{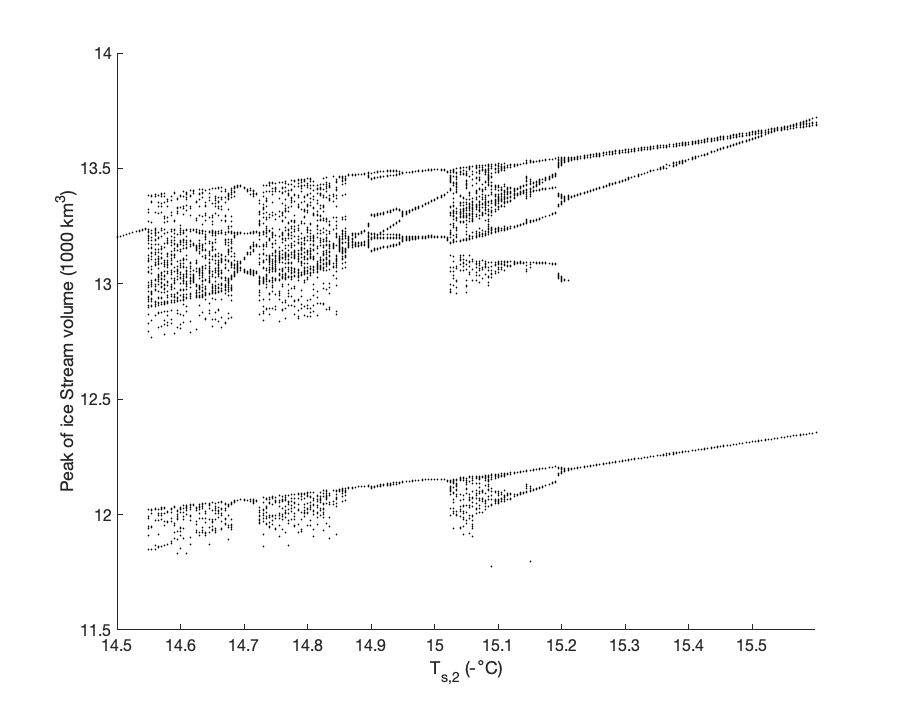} 
\caption{Bifurcation diagram for $T_{s,2}$ from 0 to 27 (top) and a zoomed view of the chaotic window between 14.6 and 15.6 (bottom). Remaining parameters are given in Tables \ref{tab:param_glob} and \ref{tab:param_box} in the Appendix.}
\label{fig:fig04}
\end{figure}

To investigate routes to chaos, we apply a parameter variation.
Figure \ref{fig:fig04} shows a bifurcation diagram, given by the peaks in the total ice volume in all three boxes, as a function of the surface temperature in B2.
It should be noted that the steady-streaming mode does not appear, as the surface temperatures of B3 is such that it is fixed in the oscillatory mode, impacting the entire system.
If all three surface temperatures were changed in unison, the steady-stream mode would be easily recovered.
There are three regimes, characterized by the rate of change of the void ratios in Boxes 1 and 2 and determined by the value of $T_{s,2}$.
These three regimes represent different dominant sources of variability.
In the first one, for $T_{s,2} < 6.65$, B2 has a large and increasing void ratio, indicating steady-streaming, and B1 has a minimal void ratio, indicating oscillatory behaviour.
As the upstream box feeds the downstream ones, the oscillations of the total volume are paced by the variability in B1.
For a large enough $T_{s,2} > 6.65 $, B2 is now in the weak build-up/surge mode, with the result being multi-mode oscillations.
For even larger $T_{s,2} > 9.05 $, the less frequent surging events allows the thickness of B1 to grow to the point where it has a positive basal melt rate and $e_1$ is constantly increasing.
That is, B2 is in the build-up/surge mode and B1 is in the steady-streaming mode, meaning B2 dictates the variability.

These regimes describe the asymptotic state, but as the void ratio can increase without bound, transitions from one regime to another can take a very long time to equilibrate depending on the initial values of these variables which may introduce long transients in the non-equilibrium setting.
In a realistic case, there might be some upper limit to the void ratio which prevents such overly long transients.

\begin{figure}[h]
        \includegraphics[width=0.8\linewidth]{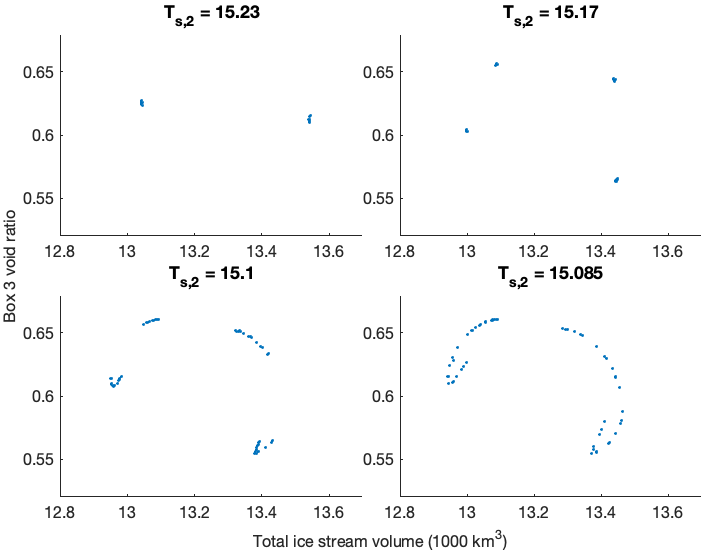}
\caption{Poincar{\'e} maps of the period-doubling route to chaos on decreasing $T_{s,2}$.}
\label{fig:fig05}
\end{figure}

The bifurcation diagram reveals a few chaotic windows.
In the parameter window $T_{s,2} \in  [14.9 , 15.4 ]$ (second panel of Fig.\ \ref{fig:fig04}), the transition from periodic to chaotic as $T_{s,2}$ is reduced occurs as a cascade of period-doubling bifurcations.
Figure \ref{fig:fig05} shows a sequence of Poincar{\'e} maps at the onset of the chaotic window.
What begins as a 2-cycle experiences a period-doubling bifurcation into a 4-cycle. 
The return map then begins to appear as a strange attractor for $T_{s,2}=15.085$. 
At the end of the chaotic window, $T_{s,2}=14.55$, the sudden transition from chaotic to periodic occurs due to the piecewise-continuous nature of the system \cite{di_bernardo_piecewise-smooth_2008}.

\begin{figure}
\includegraphics[width=0.8\linewidth]{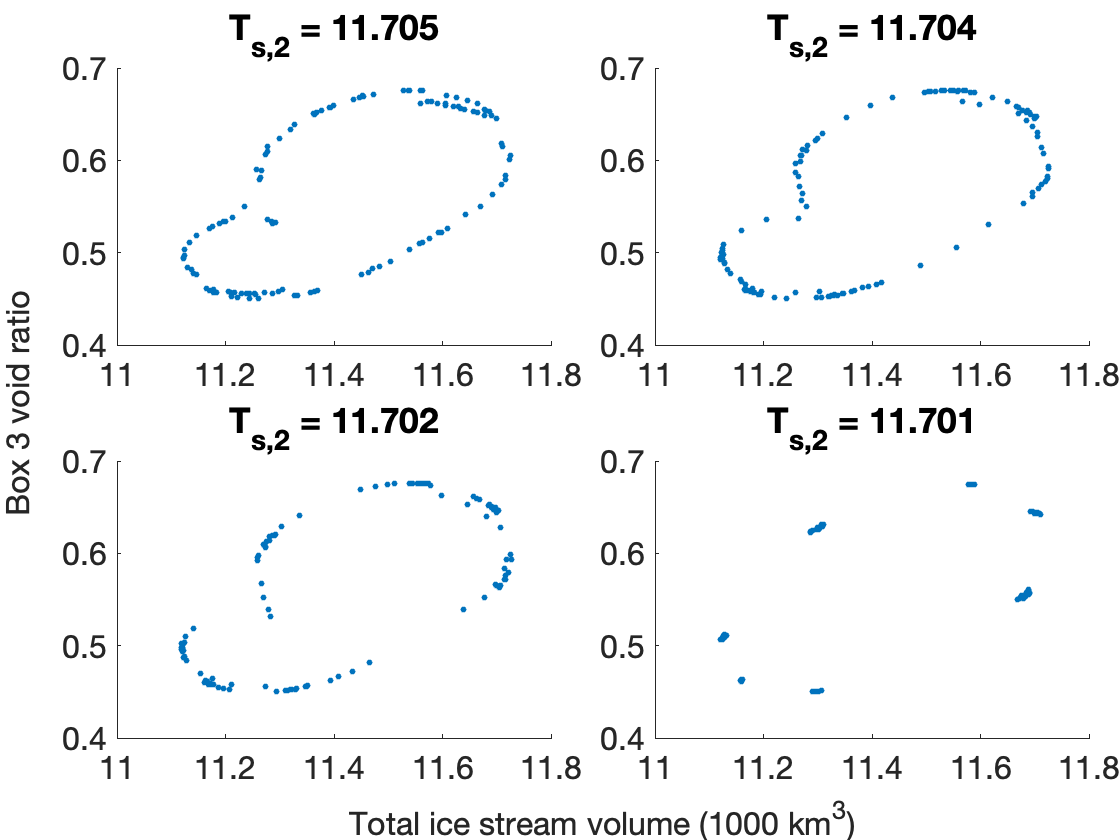}
\caption{Poincar{\'e} maps of the intermittency route to chaos on decreasing $T_{s,2}$.}
\label{fig:fig06}
\end{figure}

In a different chaotic window around $T_{s,2} =11$ in Fig. \ref{fig:fig04}, the transition back to a periodic orbit occurs as the orbits become increasingly more intermittent, with bursts of chaotic behaviour in between regular periodic motion.
This can be seen in Fig.\ \ref{fig:fig06}, which shows the same type of Poincar{\'e} maps as in Fig.\ \ref{fig:fig05}.
As $T_{s,2}$ continues to decrease, these intermittent bursts of chaos interrupting period-7 variability become more rare, as seen by the clustering of points.
This continues until the motion is purely periodic.
Hence this route to chaos, or rather route from chaos, occurs as the Pomeau-Manneville intermittency type \cite{pomeau_intermittent_1980}.

\section{Bistability, crises and chaotic transients}

In this section, the model is applied to replicate the behaviour seen in Kypke et al.\ \cite{kypke_chaotic_2025}.
In that study, there are two configurations of the ice sheet in the northwestern drainage basin with two distinct ice sheet extents.
In the `unretreated' case with a larger ice sheet extent, the ice streams oscillate with small amplitude in a more regular manner.
Under moderate warming of about 1$^{\circ}$C, the ice sheet in that region retreats and the ice streams now oscillate with larger amplitude and in a more irregular pattern.
As has been shown in the previous section, the model can shift from regularly periodic to chaotic under very small surface temperature perturbations.
We also demonstrate the transition from oscillatory to chaotic under variation of the length of one of the downstream boxes in this section to represent the retreat of the ice sheet under temperature forcing.
This comes with the caveat that the conceptual model parameters have not been altered to match those of the comprehensive ice sheet model, so the similarities are mostly qualitative. 

\begin{figure}
    	\includegraphics[width=\linewidth]{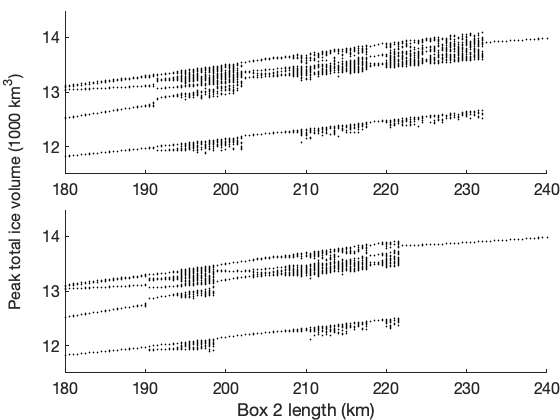}
\caption{Bifurcation diagram on varying $L_2$, displaying hysteresis of chaotic and regular periodicity. The top panel shows the case for increasing $L_2$, and the bottom panel is case for decreasing $L_2$.}
\label{fig:fig07}
\end{figure}

\begin{figure}
\includegraphics[width=1\linewidth]{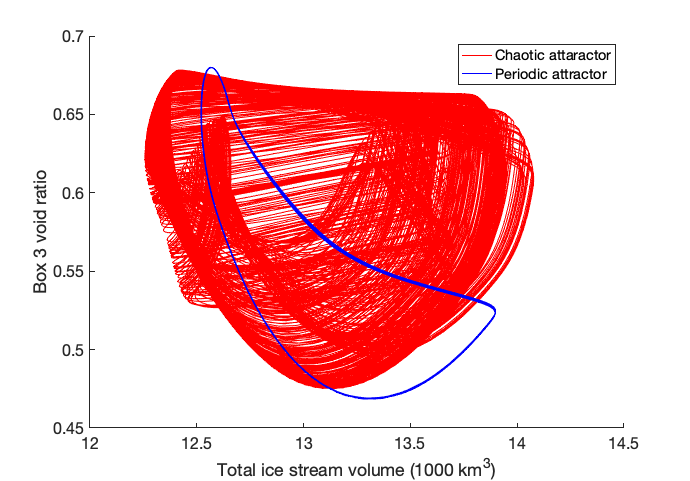}
\caption{Projection of the two attractors in the ice stream volume and B3 void ratio plane for the bistable case $L_2$ = 231.5 km, with the remaining parameters as in Tables 1 and 2 in the Appendix.}
\label{fig:fig08}
\end{figure}

Fig.~\ref{fig:fig07} shows the bifurcation diagram under variation of the length of B2 in both increasing and decreasing directions.
The region between 222 and 232 km is bistable, displaying both periodic and chaotic variability. 
At $L_2 =231.5$ km, the chaotic attractor experiences a boundary crisis \cite{grebogi_chaotic_1982}.
At $L_2 = 222$ km, the periodic attractor disappears and only the chaotic attractor remains, implying a saddle-node bifurcation of limit cycles.
The periodic motion occurs when the unfrozen till thickness of B2 ($h_{\text{till},2}$) is constant and maximal.
On the chaotic attractor, the till periodically partially refreezes. There are regions of bistability as shown, for example, in Fig.~\ref{fig:fig08}.
This explains the direction of the hysteresis: for larger ice stream lengths the ice thickness will be larger, decreasing the heat lost to conduction and maintaining a larger basal melt rate which in turn keeps the till unfrozen.
The bistability in this region also results in a fractal boundary between the basins of attraction of the chaotic and periodic attractors \cite{mcdonald_fractal_1985}.
This results in neighbouring initial conditions near the basin boundary approaching one attractor or the other, and can explain the appearance of the `anomalous' simulations in Kypke et al.\ \cite{kypke_chaotic_2025}, where trajectories forced to parameter values expected to be in the retreated configuration are in the unretreated configuration, and vice versa.
There is also another bistable region between $L_2 = 198$ and  202 km.

\begin{figure*}
\includegraphics[width=1.1\linewidth]{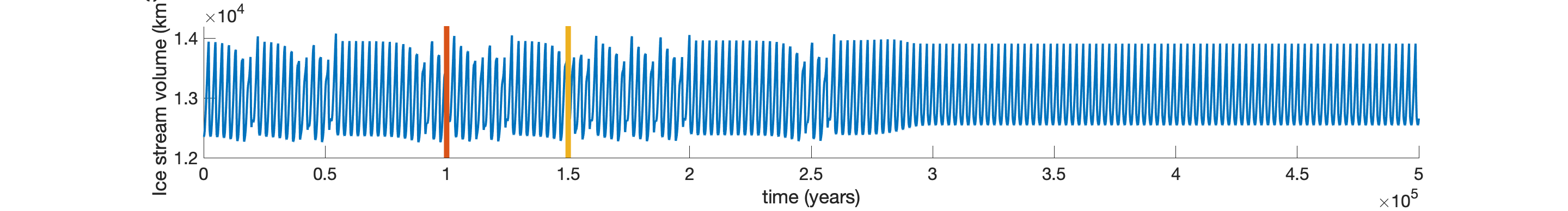}
\caption{Transient transition from chaotic to periodic for a slow parameter shift of $L_2$ from 231.5 km to 232.5 km. The orange line indicates where the parameter shift begins, and the yellow line is where it ends. The system does not transition to periodic until more than 100 kyr after the parameter has changed.}
\label{fig:fig09}
\end{figure*}

The split-stream model also displays a chaotic transient in the transition between chaotic to periodic after the boundary crisis at $L_2 = 232$ km.
Fig.~\ref{fig:fig09} shows a time series for a trajectory initialized on the chaotic attractor for $L_2 = 231.5$ km.
After 100 kyr of simulation time (orange line in Fig. \ref{fig:fig09}), the value of $L_2$ is slowly increased to 232.5 km over 50 kyr and kept constant at that value thereafter (yellow line in Fig. \ref{fig:fig09}).
The trajectory remains chaotic for a while although the system only has a periodic attractor at this parameter value.
The periodic motion then begins around 260 kyrs of simulation time, more than 100 kyr after the parameter shift is finished.
A similar chaotic transient is proposed to occur in Kypke et al. \cite{kypke_chaotic_2025}, albeit with a collapse to a state where the ice stream no longer exists rather than to a periodic orbit.

\section{Conclusions}

A model of coupled ice streams based on the simple model of R13 \cite{robel_dynamics_2013} has been constructed and its temporal dynamics investigated.
Our model displays all of the same modes of variability due to thermomechanical coupling at the base of the ice sheet: steady flow, oscillations, and build-up/surge oscillations.
Using this ice stream model it has been demonstrated that interacting ice streams can exhibit purely temporally chaotic variability over a range of parameter values.
Chaotic windows arise in a variety of ways, including period doubling, intermittency, and boundary crises.
While the coupling geometry implemented in this paper is inspired by a scenario seen in model simulations of the Greenland ice sheet, a simple reversing of the coupling direction could allow for the model to represent ice streams which converge into a common terminus.
This scenario could also be applicable to regions such as the Siple ice stream in the West Antarctic ice sheet \cite{bennett_ice_2003}.

Ice streams in realistic topographies are clearly more complicated than the idealised configuration described here. 
Their behaviour can therefore be expected to be at least as complex as in the simple model.
Since ice streams are ubiquitous to ice sheets and contribute a large amount to their overall mass balance despite their relatively small extent \cite{holland_acceleration_2008, howat_synchronous_2008, the_imbie_team_mass_2020, khan_sustained_2014, krabill_greenland_2004, larocca_greenland-wide_2023, luthcke_recent_2006, rignot_changes_2006,  rignot_acceleration_2011,  trusel_nonlinear_2018, van_den_broeke_partitioning_2009}, understanding their variability is crucial. 
That ice streams can vary chaotically means that they can sensitively depend on their initial state, complicating prospects of predicting their evolution and requiring a probabilistic framework to evaluating future mass loss. Even more surprising is that fact that there may be bistability between chaotic and non-chaotic temporal behaviour.

Various important physical processes in ice streams are not accounted for in this model. For example, a positive melt-elevation feedback which would decrease the accumulation rate at smaller ice thicknesses.
This feature is vital if one wishes to extend the model to be able to include the effects of tipping of the ice sheet that the ice streams are a part of as seen in Kypke et al. \cite{kypke_chaotic_2025}.
That is, if the ice thickness is brought low enough to decrease the accumulation rate a suitable amount, the ice stream could lose all of its mass and not recover.
The first question is whether this feedback would dominate over the feedback that allows for the oscillations.
If the surface temperature is allowed to decrease as $H$ decreases, the conductive term in the basal melt rate would be less sensitive to changes in ice thickness.
As mentioned, in R13 \cite{robel_dynamics_2013} there is no timescale associated with the vertical temperature gradient through the ice, which can occur on a timescale of thousands of years. 
The inclusion of such a propagation timescale might alleviate the complications associated with introducing an altitude-dependent accumulation rate.

\begin{acknowledgments}
This project has received funding from the European Union’s Horizon 2020 research and innovation programme under the Marie Sk{\l}odowska-Curie Innovative Training Network CriticalEarth, grant agreement no. 956170. This is a contribution to the European Union’s Horizon 2020 research and innovation programme ClimTip, grant agreement no. 101137601. 
\end{acknowledgments}

\section*{Data Availability}
The MATLAB code for running the model and plotting the figures is openly available \cite{model_code}.

\section*{Author Contribution}
{\bf K.K.}: Conceptualization, Formal analysis, Investigation, Methodology, Visualization, Writing - original draft {\bf P.A.}: Conceptualization, Formal analysis, Funding acquisition, Writing - review and editing {\bf P.D.}: Funding acquisition, Writing - review and editing.

\appendix
\section{Model parameters}

\begin{table}[ht]
\caption{\label{tab:param_glob} Shared parameter values}
\begin{ruledtabular}
\begin{tabular}{llll}
  Constant & Description & Value & Units\\
   $\rho_i$ & Density of ice  & 917 & kg m$^{-3}$ \\
   $L_f$ & Latent heat of fusion of ice & 3.335$\times 10^{5}$ & J kg$^{-1}$\\
   $K_i$ &  Thermal conductivity of ice& 2.1 &J s$^{-1}$ m$^{-1}$ K$^{-1}$   \\
   $h_b$ & Basal ice layer thickness  & 10 & m\\
   $C_{i}$ & Specific heat capacity of ice & 1.94$\times 10^{6}$ & J K$^{-1}$ m$^{-3}$\\
   $A_f $& Glen's flow law rate factor & 5$\times 10^{-25}$ & Pa$^{-3}$ s$^{-1}$\\
   $e_c$& Till consolidation void ratio  & 0.3 & ---\\
   $\tau_0$& Empirical till coefficient & 9.44$\times 10^{8}$ & Pa\\
   $c$& Empirical till exponent &21.7 & ---\\
   $q_g$& Geothermal heat flux & 0.07 & W m$^{-2}$ \\
\end{tabular}
\end{ruledtabular}
\end{table}

\begin{table}[ht]
\caption{\label{tab:param_box} Specific parameter values for Boxes  1,2 and 3 respectively}
\begin{ruledtabular}
\begin{tabular}{llll}
  Constant & Description &  Value & Units \\
  $h_{\text{till, max}}$& Maximal till thickness & 1 & m\\
  & & 2 \\
  & & 2 \\
  $a_c$& Accumulation rate & 0.05 & m yr$^{-1}$\\
    & & 0.0455&   \\
  & & 0.0417 &  \\
      $T_s$& Surface Temperature & 15 & $^{\circ}$C\\
      & &  0-30 \footnote{Parameter range in section IV. Fixed at 15 in section V} & \\
  & &   15&\\
  $L$& Ice stream length & 200& km \\
      & & 200-300 \footnote{Parameter range in section V. Fixed at 200 in section IV} &  \\
  & & 250&  \\
  $W$& Ice stream width &60 & km \\
      & & 35&  \\
  & & 25&  \\
\end{tabular}
\end{ruledtabular}
\end{table}

\newpage
\bibliography{main.bib}

\begin{thebibliography}{34}%
\makeatletter
\providecommand \@ifxundefined [1]{%
 \@ifx{#1\undefined}
}%
\providecommand \@ifnum [1]{%
 \ifnum #1\expandafter \@firstoftwo
 \else \expandafter \@secondoftwo
 \fi
}%
\providecommand \@ifx [1]{%
 \ifx #1\expandafter \@firstoftwo
 \else \expandafter \@secondoftwo
 \fi
}%
\providecommand \natexlab [1]{#1}%
\providecommand \enquote  [1]{``#1''}%
\providecommand \bibnamefont  [1]{#1}%
\providecommand \bibfnamefont [1]{#1}%
\providecommand \citenamefont [1]{#1}%
\providecommand \href@noop [0]{\@secondoftwo}%
\providecommand \href [0]{\begingroup \@sanitize@url \@href}%
\providecommand \@href[1]{\@@startlink{#1}\@@href}%
\providecommand \@@href[1]{\endgroup#1\@@endlink}%
\providecommand \@sanitize@url [0]{\catcode `\\12\catcode `\$12\catcode
  `\&12\catcode `\#12\catcode `\^12\catcode `\_12\catcode `\%12\relax}%
\providecommand \@@startlink[1]{}%
\providecommand \@@endlink[0]{}%
\providecommand \url  [0]{\begingroup\@sanitize@url \@url }%
\providecommand \@url [1]{\endgroup\@href {#1}{\urlprefix }}%
\providecommand \urlprefix  [0]{URL }%
\providecommand \Eprint [0]{\href }%
\providecommand \doibase [0]{https://doi.org/}%
\providecommand \selectlanguage [0]{\@gobble}%
\providecommand \bibinfo  [0]{\@secondoftwo}%
\providecommand \bibfield  [0]{\@secondoftwo}%
\providecommand \translation [1]{[#1]}%
\providecommand \BibitemOpen [0]{}%
\providecommand \bibitemStop [0]{}%
\providecommand \bibitemNoStop [0]{.\EOS\space}%
\providecommand \EOS [0]{\spacefactor3000\relax}%
\providecommand \BibitemShut  [1]{\csname bibitem#1\endcsname}%
\let\auto@bib@innerbib\@empty
\bibitem [{\citenamefont {Bennett}(2003)}]{bennett_ice_2003}%
  \BibitemOpen
  \bibfield  {author} {\bibinfo {author} {\bibfnamefont {M.~R.}\ \bibnamefont
  {Bennett}},\ }\bibfield  {title} {\bibinfo {title} {Ice streams as the
  arteries of an ice sheet: their mechanics, stability and significance},\
  }\href {https://doi.org/10.1016/S0012-8252(02)00130-7} {\bibfield  {journal}
  {\bibinfo  {journal} {Earth-Science Reviews}\ }\textbf {\bibinfo {volume}
  {61}},\ \bibinfo {pages} {309} (\bibinfo {year} {2003})}\BibitemShut
  {NoStop}%
\bibitem [{\citenamefont {Brinkerhoff}\ \emph {et~al.}(2017)\citenamefont
  {Brinkerhoff}, \citenamefont {Truffer},\ and\ \citenamefont
  {Aschwanden}}]{brinkerhoff_sediment_2017}%
  \BibitemOpen
  \bibfield  {author} {\bibinfo {author} {\bibfnamefont {D.}~\bibnamefont
  {Brinkerhoff}}, \bibinfo {author} {\bibfnamefont {M.}~\bibnamefont
  {Truffer}},\ and\ \bibinfo {author} {\bibfnamefont {A.}~\bibnamefont
  {Aschwanden}},\ }\bibfield  {title} {\bibinfo {title} {Sediment transport
  drives tidewater glacier periodicity},\ }\href
  {https://doi.org/10.1038/s41467-017-00095-5} {\bibfield  {journal} {\bibinfo
  {journal} {Nature Communications}\ }\textbf {\bibinfo {volume} {8}},\
  \bibinfo {pages} {90} (\bibinfo {year} {2017})}\BibitemShut {NoStop}%
\bibitem [{\citenamefont {Calov}\ \emph {et~al.}(2002)\citenamefont {Calov},
  \citenamefont {Ganopolski}, \citenamefont {Petoukhov}, \citenamefont
  {Claussen},\ and\ \citenamefont {Greve}}]{calov_largescale_2002}%
  \BibitemOpen
  \bibfield  {author} {\bibinfo {author} {\bibfnamefont {R.}~\bibnamefont
  {Calov}}, \bibinfo {author} {\bibfnamefont {A.}~\bibnamefont {Ganopolski}},
  \bibinfo {author} {\bibfnamefont {V.}~\bibnamefont {Petoukhov}}, \bibinfo
  {author} {\bibfnamefont {M.}~\bibnamefont {Claussen}},\ and\ \bibinfo
  {author} {\bibfnamefont {R.}~\bibnamefont {Greve}},\ }\bibfield  {title}
  {\bibinfo {title} {Large‐scale instabilities of the {Laurentide} ice sheet
  simulated in a fully coupled climate‐system model},\ }\bibfield  {journal}
  {\bibinfo  {journal} {Geophysical Research Letters}\ }\textbf {\bibinfo
  {volume} {29}},\ \href {https://doi.org/10.1029/2002GL016078}
  {10.1029/2002GL016078} (\bibinfo {year} {2002})\BibitemShut {NoStop}%
\bibitem [{\citenamefont {Papa}\ \emph {et~al.}(2006)\citenamefont {Papa},
  \citenamefont {Mysak},\ and\ \citenamefont {Wang}}]{papa_intermittent_2006}%
  \BibitemOpen
  \bibfield  {author} {\bibinfo {author} {\bibfnamefont {B.~D.}\ \bibnamefont
  {Papa}}, \bibinfo {author} {\bibfnamefont {L.~A.}\ \bibnamefont {Mysak}},\
  and\ \bibinfo {author} {\bibfnamefont {Z.}~\bibnamefont {Wang}},\ }\bibfield
  {title} {\bibinfo {title} {Intermittent ice sheet discharge events in
  northeastern {North} {America} during the last glacial period},\ }\href
  {https://doi.org/10.1007/s00382-005-0078-4} {\bibfield  {journal} {\bibinfo
  {journal} {Climate Dynamics}\ }\textbf {\bibinfo {volume} {26}},\ \bibinfo
  {pages} {201} (\bibinfo {year} {2006})}\BibitemShut {NoStop}%
\bibitem [{\citenamefont {Calov}\ \emph {et~al.}(2010)\citenamefont {Calov},
  \citenamefont {Greve}, \citenamefont {Abe-Ouchi}, \citenamefont {Bueler},
  \citenamefont {Huybrechts}, \citenamefont {Johnson}, \citenamefont {Pattyn},
  \citenamefont {Pollard}, \citenamefont {Ritz}, \citenamefont {Saito},\ and\
  \citenamefont {Tarasov}}]{calov_results_2010}%
  \BibitemOpen
  \bibfield  {author} {\bibinfo {author} {\bibfnamefont {R.}~\bibnamefont
  {Calov}}, \bibinfo {author} {\bibfnamefont {R.}~\bibnamefont {Greve}},
  \bibinfo {author} {\bibfnamefont {A.}~\bibnamefont {Abe-Ouchi}}, \bibinfo
  {author} {\bibfnamefont {E.}~\bibnamefont {Bueler}}, \bibinfo {author}
  {\bibfnamefont {P.}~\bibnamefont {Huybrechts}}, \bibinfo {author}
  {\bibfnamefont {J.~V.}\ \bibnamefont {Johnson}}, \bibinfo {author}
  {\bibfnamefont {F.}~\bibnamefont {Pattyn}}, \bibinfo {author} {\bibfnamefont
  {D.}~\bibnamefont {Pollard}}, \bibinfo {author} {\bibfnamefont
  {C.}~\bibnamefont {Ritz}}, \bibinfo {author} {\bibfnamefont {F.}~\bibnamefont
  {Saito}},\ and\ \bibinfo {author} {\bibfnamefont {L.}~\bibnamefont
  {Tarasov}},\ }\bibfield  {title} {\bibinfo {title} {Results from the
  {Ice}-{Sheet} {Model} {Intercomparison} {Project}–{Heinrich} {Event}
  {Intercomparison} ({ISMIP} {HEINO})},\ }\href
  {https://doi.org/10.3189/002214310792447789} {\bibfield  {journal} {\bibinfo
  {journal} {Journal of Glaciology}\ }\textbf {\bibinfo {volume} {56}},\
  \bibinfo {pages} {371} (\bibinfo {year} {2010})}\BibitemShut {NoStop}%
\bibitem [{\citenamefont {Roberts}\ \emph {et~al.}(2016)\citenamefont
  {Roberts}, \citenamefont {Payne},\ and\ \citenamefont
  {Valdes}}]{roberts_role_2016}%
  \BibitemOpen
  \bibfield  {author} {\bibinfo {author} {\bibfnamefont {W.~H.~G.}\
  \bibnamefont {Roberts}}, \bibinfo {author} {\bibfnamefont {A.~J.}\
  \bibnamefont {Payne}},\ and\ \bibinfo {author} {\bibfnamefont {P.~J.}\
  \bibnamefont {Valdes}},\ }\bibfield  {title} {\bibinfo {title} {The role of
  basal hydrology in the surging of the {Laurentide} {Ice} {Sheet}},\ }\href
  {https://doi.org/10.5194/cp-12-1601-2016} {\bibfield  {journal} {\bibinfo
  {journal} {Climate of the Past}\ }\textbf {\bibinfo {volume} {12}},\ \bibinfo
  {pages} {1601} (\bibinfo {year} {2016})}\BibitemShut {NoStop}%
\bibitem [{\citenamefont {Schannwell}\ \emph {et~al.}(2023)\citenamefont
  {Schannwell}, \citenamefont {Mikolajewicz}, \citenamefont {Ziemen},\ and\
  \citenamefont {Kapsch}}]{schannwell_sensitivity_2023}%
  \BibitemOpen
  \bibfield  {author} {\bibinfo {author} {\bibfnamefont {C.}~\bibnamefont
  {Schannwell}}, \bibinfo {author} {\bibfnamefont {U.}~\bibnamefont
  {Mikolajewicz}}, \bibinfo {author} {\bibfnamefont {F.}~\bibnamefont
  {Ziemen}},\ and\ \bibinfo {author} {\bibfnamefont {M.-L.}\ \bibnamefont
  {Kapsch}},\ }\bibfield  {title} {\bibinfo {title} {Sensitivity of
  {Heinrich}-type ice-sheet surge characteristics to boundary forcing
  perturbations},\ }\href {https://doi.org/10.5194/cp-19-179-2023} {\bibfield
  {journal} {\bibinfo  {journal} {Climate of the Past}\ }\textbf {\bibinfo
  {volume} {19}},\ \bibinfo {pages} {179} (\bibinfo {year} {2023})}\BibitemShut
  {NoStop}%
\bibitem [{\citenamefont {Heinrich}(1988)}]{heinrich_origin_1988}%
  \BibitemOpen
  \bibfield  {author} {\bibinfo {author} {\bibfnamefont {H.}~\bibnamefont
  {Heinrich}},\ }\bibfield  {title} {\bibinfo {title} {Origin and
  {Consequences} of {Cyclic} {Ice} {Rafting} in the {Northeast} {Atlantic}
  {Ocean} {During} the {Past} 130,000 {Years}},\ }\href
  {https://doi.org/10.1016/0033-5894(88)90057-9} {\bibfield  {journal}
  {\bibinfo  {journal} {Quaternary Research}\ }\textbf {\bibinfo {volume}
  {29}},\ \bibinfo {pages} {142} (\bibinfo {year} {1988})}\BibitemShut
  {NoStop}%
\bibitem [{\citenamefont {Broecker}\ \emph {et~al.}(1992)\citenamefont
  {Broecker}, \citenamefont {Bond}, \citenamefont {Klas}, \citenamefont
  {Clark},\ and\ \citenamefont {McManus}}]{broecker_origin_1992}%
  \BibitemOpen
  \bibfield  {author} {\bibinfo {author} {\bibfnamefont {W.}~\bibnamefont
  {Broecker}}, \bibinfo {author} {\bibfnamefont {G.}~\bibnamefont {Bond}},
  \bibinfo {author} {\bibfnamefont {M.}~\bibnamefont {Klas}}, \bibinfo {author}
  {\bibfnamefont {E.}~\bibnamefont {Clark}},\ and\ \bibinfo {author}
  {\bibfnamefont {J.}~\bibnamefont {McManus}},\ }\bibfield  {title} {\bibinfo
  {title} {Origin of the northern {Atlantic}'s {Heinrich} events},\ }\href
  {https://doi.org/10.1007/BF00193540} {\bibfield  {journal} {\bibinfo
  {journal} {Climate Dynamics}\ }\textbf {\bibinfo {volume} {6}},\ \bibinfo
  {pages} {265} (\bibinfo {year} {1992})}\BibitemShut {NoStop}%
\bibitem [{\citenamefont {Dansgaard}\ \emph {et~al.}(1993)\citenamefont
  {Dansgaard}, \citenamefont {Johnsen}, \citenamefont {Clausen}, \citenamefont
  {Dahl-Jensen}, \citenamefont {Gundestrup}, \citenamefont {Hammer},
  \citenamefont {Hvidberg}, \citenamefont {Steffensen}, \citenamefont
  {Sveinbjörnsdottir}, \citenamefont {Jouzel},\ and\ \citenamefont
  {Bond}}]{dansgaard_evidence_1993}%
  \BibitemOpen
  \bibfield  {author} {\bibinfo {author} {\bibfnamefont {W.}~\bibnamefont
  {Dansgaard}}, \bibinfo {author} {\bibfnamefont {S.~J.}\ \bibnamefont
  {Johnsen}}, \bibinfo {author} {\bibfnamefont {H.~B.}\ \bibnamefont
  {Clausen}}, \bibinfo {author} {\bibfnamefont {D.}~\bibnamefont
  {Dahl-Jensen}}, \bibinfo {author} {\bibfnamefont {N.~S.}\ \bibnamefont
  {Gundestrup}}, \bibinfo {author} {\bibfnamefont {C.~U.}\ \bibnamefont
  {Hammer}}, \bibinfo {author} {\bibfnamefont {C.~S.}\ \bibnamefont
  {Hvidberg}}, \bibinfo {author} {\bibfnamefont {J.~P.}\ \bibnamefont
  {Steffensen}}, \bibinfo {author} {\bibfnamefont {A.~E.}\ \bibnamefont
  {Sveinbjörnsdottir}}, \bibinfo {author} {\bibfnamefont {J.}~\bibnamefont
  {Jouzel}},\ and\ \bibinfo {author} {\bibfnamefont {G.}~\bibnamefont {Bond}},\
  }\bibfield  {title} {\bibinfo {title} {Evidence for general instability of
  past climate from a 250-kyr ice-core record},\ }\href
  {https://doi.org/10.1038/364218a0} {\bibfield  {journal} {\bibinfo  {journal}
  {Nature}\ }\textbf {\bibinfo {volume} {364}},\ \bibinfo {pages} {218}
  (\bibinfo {year} {1993})}\BibitemShut {NoStop}%
\bibitem [{\citenamefont {Alvarez-Solas}\ \emph {et~al.}(2013)\citenamefont
  {Alvarez-Solas}, \citenamefont {Robinson}, \citenamefont {Montoya},\ and\
  \citenamefont {Ritz}}]{alvarez-solas_iceberg_2013}%
  \BibitemOpen
  \bibfield  {author} {\bibinfo {author} {\bibfnamefont {J.}~\bibnamefont
  {Alvarez-Solas}}, \bibinfo {author} {\bibfnamefont {A.}~\bibnamefont
  {Robinson}}, \bibinfo {author} {\bibfnamefont {M.}~\bibnamefont {Montoya}},\
  and\ \bibinfo {author} {\bibfnamefont {C.}~\bibnamefont {Ritz}},\ }\bibfield
  {title} {\bibinfo {title} {Iceberg discharges of the last glacial period
  driven by oceanic circulation changes},\ }\href
  {https://doi.org/10.1073/pnas.1306622110} {\bibfield  {journal} {\bibinfo
  {journal} {Proceedings of the National Academy of Sciences}\ }\textbf
  {\bibinfo {volume} {110}},\ \bibinfo {pages} {16350} (\bibinfo {year}
  {2013})}\BibitemShut {NoStop}%
\bibitem [{\citenamefont {Payne}(1995)}]{payne_limit_1995}%
  \BibitemOpen
  \bibfield  {author} {\bibinfo {author} {\bibfnamefont {A.~J.}\ \bibnamefont
  {Payne}},\ }\bibfield  {title} {\bibinfo {title} {Limit cycles in the basal
  thermal regime of ice sheets},\ }\href {https://doi.org/10.1029/94JB02778}
  {\bibfield  {journal} {\bibinfo  {journal} {Journal of Geophysical Research:
  Solid Earth}\ }\textbf {\bibinfo {volume} {100}},\ \bibinfo {pages} {4249}
  (\bibinfo {year} {1995})}\BibitemShut {NoStop}%
\bibitem [{\citenamefont {Robel}\ \emph {et~al.}(2013)\citenamefont {Robel},
  \citenamefont {DeGiuli}, \citenamefont {Schoof},\ and\ \citenamefont
  {Tziperman}}]{robel_dynamics_2013}%
  \BibitemOpen
  \bibfield  {author} {\bibinfo {author} {\bibfnamefont {A.~A.}\ \bibnamefont
  {Robel}}, \bibinfo {author} {\bibfnamefont {E.}~\bibnamefont {DeGiuli}},
  \bibinfo {author} {\bibfnamefont {C.}~\bibnamefont {Schoof}},\ and\ \bibinfo
  {author} {\bibfnamefont {E.}~\bibnamefont {Tziperman}},\ }\bibfield  {title}
  {\bibinfo {title} {Dynamics of ice stream temporal variability: {Modes},
  scales, and hysteresis},\ }\href {https://doi.org/10.1002/jgrf.20072}
  {\bibfield  {journal} {\bibinfo  {journal} {Journal of Geophysical Research:
  Earth Surface}\ }\textbf {\bibinfo {volume} {118}},\ \bibinfo {pages} {925}
  (\bibinfo {year} {2013})}\BibitemShut {NoStop}%
\bibitem [{\citenamefont {Mantelli}\ \emph {et~al.}(2016)\citenamefont
  {Mantelli}, \citenamefont {Bertagni},\ and\ \citenamefont
  {Ridolfi}}]{mantelli_stochastic_2016}%
  \BibitemOpen
  \bibfield  {author} {\bibinfo {author} {\bibfnamefont {E.}~\bibnamefont
  {Mantelli}}, \bibinfo {author} {\bibfnamefont {M.~B.}\ \bibnamefont
  {Bertagni}},\ and\ \bibinfo {author} {\bibfnamefont {L.}~\bibnamefont
  {Ridolfi}},\ }\bibfield  {title} {\bibinfo {title} {Stochastic ice stream
  dynamics},\ }\bibfield  {journal} {\bibinfo  {journal} {Proceedings of the
  National Academy of Sciences}\ }\textbf {\bibinfo {volume} {113}},\ \href
  {https://doi.org/10.1073/pnas.1600362113} {10.1073/pnas.1600362113} (\bibinfo
  {year} {2016})\BibitemShut {NoStop}%
\bibitem [{\citenamefont {Kypke}\ \emph
  {et~al.}(2025{\natexlab{a}})\citenamefont {Kypke}, \citenamefont {Montoya},
  \citenamefont {Robinson}, \citenamefont {Alvarez-Solas}, \citenamefont
  {Swierczek-Jereczek},\ and\ \citenamefont {Ditlevsen}}]{kypke_chaotic_2025}%
  \BibitemOpen
  \bibfield  {author} {\bibinfo {author} {\bibfnamefont {K.}~\bibnamefont
  {Kypke}}, \bibinfo {author} {\bibfnamefont {M.}~\bibnamefont {Montoya}},
  \bibinfo {author} {\bibfnamefont {A.}~\bibnamefont {Robinson}}, \bibinfo
  {author} {\bibfnamefont {J.}~\bibnamefont {Alvarez-Solas}}, \bibinfo {author}
  {\bibfnamefont {J.}~\bibnamefont {Swierczek-Jereczek}},\ and\ \bibinfo
  {author} {\bibfnamefont {P.}~\bibnamefont {Ditlevsen}},\ }\bibfield  {title}
  {\bibinfo {title} {Chaotic fluctuations in {G}reenland outlet glaciers limit
  predictability of a future ice sheet collapse},\ }\href
  {https://doi.org/10.5194/egusphere-2025-4116} {\bibfield  {journal} {\bibinfo
   {journal} {EGUsphere}\ }\textbf {\bibinfo {volume} {2025}},\ \bibinfo
  {pages} {1} (\bibinfo {year} {2025}{\natexlab{a}})}\BibitemShut {NoStop}%
\bibitem [{\citenamefont {Lai}\ and\ \citenamefont
  {Tél}(2011)}]{lai_transient_2011}%
  \BibitemOpen
  \bibfield  {author} {\bibinfo {author} {\bibfnamefont {Y.-C.}\ \bibnamefont
  {Lai}}\ and\ \bibinfo {author} {\bibfnamefont {T.}~\bibnamefont {Tél}},\
  }\href {https://doi.org/10.1007/978-1-4419-6987-3} {\emph {\bibinfo {title}
  {Transient {Chaos}}}},\ \bibinfo {series} {Applied {Mathematical}
  {Sciences}}, Vol.\ \bibinfo {volume} {173}\ (\bibinfo  {publisher} {Springer
  New York},\ \bibinfo {address} {New York, NY},\ \bibinfo {year}
  {2011})\BibitemShut {NoStop}%
\bibitem [{\citenamefont {Tulaczyk}\ \emph {et~al.}(2000)\citenamefont
  {Tulaczyk}, \citenamefont {Kamb},\ and\ \citenamefont
  {Engelhardt}}]{tulaczyk_basal_2000}%
  \BibitemOpen
  \bibfield  {author} {\bibinfo {author} {\bibfnamefont {S.}~\bibnamefont
  {Tulaczyk}}, \bibinfo {author} {\bibfnamefont {W.~B.}\ \bibnamefont {Kamb}},\
  and\ \bibinfo {author} {\bibfnamefont {H.~F.}\ \bibnamefont {Engelhardt}},\
  }\bibfield  {title} {\bibinfo {title} {Basal mechanics of {Ice} {Stream} {B},
  west {Antarctica}: 1. {Till} mechanics},\ }\href
  {https://doi.org/10.1029/1999JB900329} {\bibfield  {journal} {\bibinfo
  {journal} {Journal of Geophysical Research: Solid Earth}\ }\textbf {\bibinfo
  {volume} {105}},\ \bibinfo {pages} {463} (\bibinfo {year}
  {2000})}\BibitemShut {NoStop}%
\bibitem [{\citenamefont {Paterson}(1994)}]{paterson_structure_1994}%
  \BibitemOpen
  \bibfield  {author} {\bibinfo {author} {\bibfnamefont {W.}~\bibnamefont
  {Paterson}},\ }\bibfield  {title} {\bibinfo {title} {Structure and
  {Deformation} of {Ice}},\ }in\ \href
  {https://doi.org/10.1016/B978-0-08-037944-9.50011-X} {\emph {\bibinfo
  {booktitle} {The {Physics} of {Glaciers}}}}\ (\bibinfo  {publisher}
  {Elsevier},\ \bibinfo {year} {1994})\ pp.\ \bibinfo {pages}
  {78--102}\BibitemShut {NoStop}%
\bibitem [{\citenamefont
  {Di~Bernardo}(2008)}]{di_bernardo_piecewise-smooth_2008}%
  \BibitemOpen
  \bibinfo {editor} {\bibfnamefont {M.}~\bibnamefont {Di~Bernardo}},\ ed.,\
  \href@noop {} {\emph {\bibinfo {title} {Piecewise-smooth dynamical systems:
  theory and applications}}},\ \bibinfo {series} {Applied mathematical
  sciences}\ No.\ \bibinfo {number} {163}\ (\bibinfo  {publisher} {Springer},\
  \bibinfo {address} {London},\ \bibinfo {year} {2008})\BibitemShut {NoStop}%
\bibitem [{\citenamefont {Pomeau}\ and\ \citenamefont
  {Manneville}(1980)}]{pomeau_intermittent_1980}%
  \BibitemOpen
  \bibfield  {author} {\bibinfo {author} {\bibfnamefont {Y.}~\bibnamefont
  {Pomeau}}\ and\ \bibinfo {author} {\bibfnamefont {P.}~\bibnamefont
  {Manneville}},\ }\bibfield  {title} {\bibinfo {title} {Intermittent
  transition to turbulence in dissipative dynamical systems},\ }\href
  {https://doi.org/10.1007/BF01197757} {\bibfield  {journal} {\bibinfo
  {journal} {Communications in Mathematical Physics}\ }\textbf {\bibinfo
  {volume} {74}},\ \bibinfo {pages} {189} (\bibinfo {year} {1980})}\BibitemShut
  {NoStop}%
\bibitem [{\citenamefont {Grebogi}\ \emph {et~al.}(1982)\citenamefont
  {Grebogi}, \citenamefont {Ott},\ and\ \citenamefont
  {Yorke}}]{grebogi_chaotic_1982}%
  \BibitemOpen
  \bibfield  {author} {\bibinfo {author} {\bibfnamefont {C.}~\bibnamefont
  {Grebogi}}, \bibinfo {author} {\bibfnamefont {E.}~\bibnamefont {Ott}},\ and\
  \bibinfo {author} {\bibfnamefont {J.~A.}\ \bibnamefont {Yorke}},\ }\bibfield
  {title} {\bibinfo {title} {Chaotic {Attractors} in {Crisis}},\ }\href
  {https://doi.org/10.1103/PhysRevLett.48.1507} {\bibfield  {journal} {\bibinfo
   {journal} {Physical Review Letters}\ }\textbf {\bibinfo {volume} {48}},\
  \bibinfo {pages} {1507} (\bibinfo {year} {1982})}\BibitemShut {NoStop}%
\bibitem [{\citenamefont {McDonald}\ \emph {et~al.}(1985)\citenamefont
  {McDonald}, \citenamefont {Grebogi}, \citenamefont {Ott},\ and\ \citenamefont
  {Yorke}}]{mcdonald_fractal_1985}%
  \BibitemOpen
  \bibfield  {author} {\bibinfo {author} {\bibfnamefont {S.~W.}\ \bibnamefont
  {McDonald}}, \bibinfo {author} {\bibfnamefont {C.}~\bibnamefont {Grebogi}},
  \bibinfo {author} {\bibfnamefont {E.}~\bibnamefont {Ott}},\ and\ \bibinfo
  {author} {\bibfnamefont {J.~A.}\ \bibnamefont {Yorke}},\ }\bibfield  {title}
  {\bibinfo {title} {Fractal basin boundaries},\ }\href
  {https://doi.org/10.1016/0167-2789(85)90001-6} {\bibfield  {journal}
  {\bibinfo  {journal} {Physica D: Nonlinear Phenomena}\ }\textbf {\bibinfo
  {volume} {17}},\ \bibinfo {pages} {125} (\bibinfo {year} {1985})}\BibitemShut
  {NoStop}%
\bibitem [{\citenamefont {Holland}\ \emph {et~al.}(2008)\citenamefont
  {Holland}, \citenamefont {Thomas}, \citenamefont {de~Young}, \citenamefont
  {Ribergaard},\ and\ \citenamefont {Lyberth}}]{holland_acceleration_2008}%
  \BibitemOpen
  \bibfield  {author} {\bibinfo {author} {\bibfnamefont {D.~M.}\ \bibnamefont
  {Holland}}, \bibinfo {author} {\bibfnamefont {R.~H.}\ \bibnamefont {Thomas}},
  \bibinfo {author} {\bibfnamefont {B.}~\bibnamefont {de~Young}}, \bibinfo
  {author} {\bibfnamefont {M.~H.}\ \bibnamefont {Ribergaard}},\ and\ \bibinfo
  {author} {\bibfnamefont {B.}~\bibnamefont {Lyberth}},\ }\bibfield  {title}
  {\bibinfo {title} {Acceleration of {Jakobshavn} {Isbræ} triggered by warm
  subsurface ocean waters},\ }\href {https://doi.org/10.1038/ngeo316}
  {\bibfield  {journal} {\bibinfo  {journal} {Nature Geoscience}\ }\textbf
  {\bibinfo {volume} {1}},\ \bibinfo {pages} {659} (\bibinfo {year}
  {2008})}\BibitemShut {NoStop}%
\bibitem [{\citenamefont {Howat}\ \emph {et~al.}(2008)\citenamefont {Howat},
  \citenamefont {Joughin}, \citenamefont {Fahnestock}, \citenamefont {Smith},\
  and\ \citenamefont {Scambos}}]{howat_synchronous_2008}%
  \BibitemOpen
  \bibfield  {author} {\bibinfo {author} {\bibfnamefont {I.~M.}\ \bibnamefont
  {Howat}}, \bibinfo {author} {\bibfnamefont {I.}~\bibnamefont {Joughin}},
  \bibinfo {author} {\bibfnamefont {M.}~\bibnamefont {Fahnestock}}, \bibinfo
  {author} {\bibfnamefont {B.~E.}\ \bibnamefont {Smith}},\ and\ \bibinfo
  {author} {\bibfnamefont {T.~A.}\ \bibnamefont {Scambos}},\ }\bibfield
  {title} {\bibinfo {title} {Synchronous retreat and acceleration of southeast
  {Greenland} outlet glaciers 2000–06: ice dynamics and coupling to
  climate},\ }\href {https://doi.org/10.3189/002214308786570908} {\bibfield
  {journal} {\bibinfo  {journal} {Journal of Glaciology}\ }\textbf {\bibinfo
  {volume} {54}},\ \bibinfo {pages} {646} (\bibinfo {year} {2008})}\BibitemShut
  {NoStop}%
\bibitem [{\citenamefont {{The IMBIE Team}}(2020)}]{the_imbie_team_mass_2020}%
  \BibitemOpen
  \bibfield  {author} {\bibinfo {author} {\bibnamefont {{The IMBIE Team}}},\
  }\bibfield  {title} {\bibinfo {title} {Mass balance of the {Greenland} {Ice}
  {Sheet} from 1992 to 2018},\ }\href
  {https://doi.org/10.1038/s41586-019-1855-2} {\bibfield  {journal} {\bibinfo
  {journal} {Nature}\ }\textbf {\bibinfo {volume} {579}},\ \bibinfo {pages}
  {233} (\bibinfo {year} {2020})}\BibitemShut {NoStop}%
\bibitem [{\citenamefont {Khan}\ \emph {et~al.}(2014)\citenamefont {Khan},
  \citenamefont {Kjær}, \citenamefont {Bevis}, \citenamefont {Bamber},
  \citenamefont {Wahr}, \citenamefont {Kjeldsen}, \citenamefont {Bjørk},
  \citenamefont {Korsgaard}, \citenamefont {Stearns}, \citenamefont {Van
  Den~Broeke}, \citenamefont {Liu}, \citenamefont {Larsen},\ and\ \citenamefont
  {Muresan}}]{khan_sustained_2014}%
  \BibitemOpen
  \bibfield  {author} {\bibinfo {author} {\bibfnamefont {S.~A.}\ \bibnamefont
  {Khan}}, \bibinfo {author} {\bibfnamefont {K.~H.}\ \bibnamefont {Kjær}},
  \bibinfo {author} {\bibfnamefont {M.}~\bibnamefont {Bevis}}, \bibinfo
  {author} {\bibfnamefont {J.~L.}\ \bibnamefont {Bamber}}, \bibinfo {author}
  {\bibfnamefont {J.}~\bibnamefont {Wahr}}, \bibinfo {author} {\bibfnamefont
  {K.~K.}\ \bibnamefont {Kjeldsen}}, \bibinfo {author} {\bibfnamefont {A.~A.}\
  \bibnamefont {Bjørk}}, \bibinfo {author} {\bibfnamefont {N.~J.}\
  \bibnamefont {Korsgaard}}, \bibinfo {author} {\bibfnamefont {L.~A.}\
  \bibnamefont {Stearns}}, \bibinfo {author} {\bibfnamefont {M.~R.}\
  \bibnamefont {Van Den~Broeke}}, \bibinfo {author} {\bibfnamefont
  {L.}~\bibnamefont {Liu}}, \bibinfo {author} {\bibfnamefont {N.~K.}\
  \bibnamefont {Larsen}},\ and\ \bibinfo {author} {\bibfnamefont {I.~S.}\
  \bibnamefont {Muresan}},\ }\bibfield  {title} {\bibinfo {title} {Sustained
  mass loss of the northeast {Greenland} ice sheet triggered by regional
  warming},\ }\href {https://doi.org/10.1038/nclimate2161} {\bibfield
  {journal} {\bibinfo  {journal} {Nature Climate Change}\ }\textbf {\bibinfo
  {volume} {4}},\ \bibinfo {pages} {292} (\bibinfo {year} {2014})}\BibitemShut
  {NoStop}%
\bibitem [{\citenamefont {Krabill}\ \emph {et~al.}(2004)\citenamefont
  {Krabill}, \citenamefont {Hanna}, \citenamefont {Huybrechts}, \citenamefont
  {Abdalati}, \citenamefont {Cappelen}, \citenamefont {Csatho}, \citenamefont
  {Frederick}, \citenamefont {Manizade}, \citenamefont {Martin}, \citenamefont
  {Sonntag}, \citenamefont {Swift}, \citenamefont {Thomas},\ and\ \citenamefont
  {Yungel}}]{krabill_greenland_2004}%
  \BibitemOpen
  \bibfield  {author} {\bibinfo {author} {\bibfnamefont {W.}~\bibnamefont
  {Krabill}}, \bibinfo {author} {\bibfnamefont {E.}~\bibnamefont {Hanna}},
  \bibinfo {author} {\bibfnamefont {P.}~\bibnamefont {Huybrechts}}, \bibinfo
  {author} {\bibfnamefont {W.}~\bibnamefont {Abdalati}}, \bibinfo {author}
  {\bibfnamefont {J.}~\bibnamefont {Cappelen}}, \bibinfo {author}
  {\bibfnamefont {B.}~\bibnamefont {Csatho}}, \bibinfo {author} {\bibfnamefont
  {E.}~\bibnamefont {Frederick}}, \bibinfo {author} {\bibfnamefont
  {S.}~\bibnamefont {Manizade}}, \bibinfo {author} {\bibfnamefont
  {C.}~\bibnamefont {Martin}}, \bibinfo {author} {\bibfnamefont
  {J.}~\bibnamefont {Sonntag}}, \bibinfo {author} {\bibfnamefont
  {R.}~\bibnamefont {Swift}}, \bibinfo {author} {\bibfnamefont
  {R.}~\bibnamefont {Thomas}},\ and\ \bibinfo {author} {\bibfnamefont
  {J.}~\bibnamefont {Yungel}},\ }\bibfield  {title} {\bibinfo {title}
  {Greenland {Ice} {Sheet}: {Increased} coastal thinning},\ }\bibfield
  {journal} {\bibinfo  {journal} {Geophysical Research Letters}\ }\textbf
  {\bibinfo {volume} {31}},\ \href {https://doi.org/10.1029/2004GL021533}
  {10.1029/2004GL021533} (\bibinfo {year} {2004})\BibitemShut {NoStop}%
\bibitem [{\citenamefont {Larocca}\ \emph {et~al.}(2023)\citenamefont
  {Larocca}, \citenamefont {Twining–Ward}, \citenamefont {Axford},
  \citenamefont {Schweinsberg}, \citenamefont {Larsen}, \citenamefont
  {Westergaard–Nielsen}, \citenamefont {Luetzenburg}, \citenamefont {Briner},
  \citenamefont {Kjeldsen},\ and\ \citenamefont
  {Bjørk}}]{larocca_greenland-wide_2023}%
  \BibitemOpen
  \bibfield  {author} {\bibinfo {author} {\bibfnamefont {L.~J.}\ \bibnamefont
  {Larocca}}, \bibinfo {author} {\bibfnamefont {M.}~\bibnamefont
  {Twining–Ward}}, \bibinfo {author} {\bibfnamefont {Y.}~\bibnamefont
  {Axford}}, \bibinfo {author} {\bibfnamefont {A.~D.}\ \bibnamefont
  {Schweinsberg}}, \bibinfo {author} {\bibfnamefont {S.~H.}\ \bibnamefont
  {Larsen}}, \bibinfo {author} {\bibfnamefont {A.}~\bibnamefont
  {Westergaard–Nielsen}}, \bibinfo {author} {\bibfnamefont {G.}~\bibnamefont
  {Luetzenburg}}, \bibinfo {author} {\bibfnamefont {J.~P.}\ \bibnamefont
  {Briner}}, \bibinfo {author} {\bibfnamefont {K.~K.}\ \bibnamefont
  {Kjeldsen}},\ and\ \bibinfo {author} {\bibfnamefont {A.~A.}\ \bibnamefont
  {Bjørk}},\ }\bibfield  {title} {\bibinfo {title} {Greenland-wide accelerated
  retreat of peripheral glaciers in the twenty-first century},\ }\href
  {https://doi.org/10.1038/s41558-023-01855-6} {\bibfield  {journal} {\bibinfo
  {journal} {Nature Climate Change}\ ,\ \bibinfo {pages} {1}} (\bibinfo {year}
  {2023})}\BibitemShut {NoStop}%
\bibitem [{\citenamefont {Luthcke}\ \emph {et~al.}(2006)\citenamefont
  {Luthcke}, \citenamefont {Zwally}, \citenamefont {Abdalati}, \citenamefont
  {Rowlands}, \citenamefont {Ray}, \citenamefont {Nerem}, \citenamefont
  {Lemoine}, \citenamefont {McCarthy},\ and\ \citenamefont
  {Chinn}}]{luthcke_recent_2006}%
  \BibitemOpen
  \bibfield  {author} {\bibinfo {author} {\bibfnamefont {S.~B.}\ \bibnamefont
  {Luthcke}}, \bibinfo {author} {\bibfnamefont {H.~J.}\ \bibnamefont {Zwally}},
  \bibinfo {author} {\bibfnamefont {W.}~\bibnamefont {Abdalati}}, \bibinfo
  {author} {\bibfnamefont {D.~D.}\ \bibnamefont {Rowlands}}, \bibinfo {author}
  {\bibfnamefont {R.~D.}\ \bibnamefont {Ray}}, \bibinfo {author} {\bibfnamefont
  {R.~S.}\ \bibnamefont {Nerem}}, \bibinfo {author} {\bibfnamefont {F.~G.}\
  \bibnamefont {Lemoine}}, \bibinfo {author} {\bibfnamefont {J.~J.}\
  \bibnamefont {McCarthy}},\ and\ \bibinfo {author} {\bibfnamefont {D.~S.}\
  \bibnamefont {Chinn}},\ }\bibfield  {title} {\bibinfo {title} {Recent
  {Greenland} {Ice} {Mass} {Loss} by {Drainage} {System} from {Satellite}
  {Gravity} {Observations}},\ }\href {https://doi.org/10.1126/science.1130776}
  {\bibfield  {journal} {\bibinfo  {journal} {Science}\ }\textbf {\bibinfo
  {volume} {314}},\ \bibinfo {pages} {1286} (\bibinfo {year}
  {2006})}\BibitemShut {NoStop}%
\bibitem [{\citenamefont {Rignot}\ and\ \citenamefont
  {Kanagaratnam}(2006)}]{rignot_changes_2006}%
  \BibitemOpen
  \bibfield  {author} {\bibinfo {author} {\bibfnamefont {E.}~\bibnamefont
  {Rignot}}\ and\ \bibinfo {author} {\bibfnamefont {P.}~\bibnamefont
  {Kanagaratnam}},\ }\bibfield  {title} {\bibinfo {title} {Changes in the
  {Velocity} {Structure} of the {Greenland} {Ice} {Sheet}},\ }\href
  {https://doi.org/10.1126/science.1121381} {\bibfield  {journal} {\bibinfo
  {journal} {Science}\ }\textbf {\bibinfo {volume} {311}},\ \bibinfo {pages}
  {986} (\bibinfo {year} {2006})}\BibitemShut {NoStop}%
\bibitem [{\citenamefont {Rignot}\ \emph {et~al.}(2011)\citenamefont {Rignot},
  \citenamefont {Velicogna}, \citenamefont {van~den Broeke}, \citenamefont
  {Monaghan},\ and\ \citenamefont {Lenaerts}}]{rignot_acceleration_2011}%
  \BibitemOpen
  \bibfield  {author} {\bibinfo {author} {\bibfnamefont {E.}~\bibnamefont
  {Rignot}}, \bibinfo {author} {\bibfnamefont {I.}~\bibnamefont {Velicogna}},
  \bibinfo {author} {\bibfnamefont {M.~R.}\ \bibnamefont {van~den Broeke}},
  \bibinfo {author} {\bibfnamefont {A.}~\bibnamefont {Monaghan}},\ and\
  \bibinfo {author} {\bibfnamefont {J.~T.~M.}\ \bibnamefont {Lenaerts}},\
  }\bibfield  {title} {\bibinfo {title} {Acceleration of the contribution of
  the {Greenland} and {Antarctic} ice sheets to sea level rise},\ }\bibfield
  {journal} {\bibinfo  {journal} {Geophysical Research Letters}\ }\textbf
  {\bibinfo {volume} {38}},\ \href {https://doi.org/10.1029/2011GL046583}
  {10.1029/2011GL046583} (\bibinfo {year} {2011})\BibitemShut {NoStop}%
\bibitem [{\citenamefont {Trusel}\ \emph {et~al.}(2018)\citenamefont {Trusel},
  \citenamefont {Das}, \citenamefont {Osman}, \citenamefont {Evans},
  \citenamefont {Smith}, \citenamefont {Fettweis}, \citenamefont {McConnell},
  \citenamefont {Noël},\ and\ \citenamefont {Van
  Den~Broeke}}]{trusel_nonlinear_2018}%
  \BibitemOpen
  \bibfield  {author} {\bibinfo {author} {\bibfnamefont {L.~D.}\ \bibnamefont
  {Trusel}}, \bibinfo {author} {\bibfnamefont {S.~B.}\ \bibnamefont {Das}},
  \bibinfo {author} {\bibfnamefont {M.~B.}\ \bibnamefont {Osman}}, \bibinfo
  {author} {\bibfnamefont {M.~J.}\ \bibnamefont {Evans}}, \bibinfo {author}
  {\bibfnamefont {B.~E.}\ \bibnamefont {Smith}}, \bibinfo {author}
  {\bibfnamefont {X.}~\bibnamefont {Fettweis}}, \bibinfo {author}
  {\bibfnamefont {J.~R.}\ \bibnamefont {McConnell}}, \bibinfo {author}
  {\bibfnamefont {B.~P.~Y.}\ \bibnamefont {Noël}},\ and\ \bibinfo {author}
  {\bibfnamefont {M.~R.}\ \bibnamefont {Van Den~Broeke}},\ }\bibfield  {title}
  {\bibinfo {title} {Nonlinear rise in {Greenland} runoff in response to
  post-industrial {Arctic} warming},\ }\href
  {https://doi.org/10.1038/s41586-018-0752-4} {\bibfield  {journal} {\bibinfo
  {journal} {Nature}\ }\textbf {\bibinfo {volume} {564}},\ \bibinfo {pages}
  {104} (\bibinfo {year} {2018})}\BibitemShut {NoStop}%
\bibitem [{\citenamefont {Van Den~Broeke}\ \emph {et~al.}(2009)\citenamefont
  {Van Den~Broeke}, \citenamefont {Bamber}, \citenamefont {Ettema},
  \citenamefont {Rignot}, \citenamefont {Schrama}, \citenamefont {Van De~Berg},
  \citenamefont {Van~Meijgaard}, \citenamefont {Velicogna},\ and\ \citenamefont
  {Wouters}}]{van_den_broeke_partitioning_2009}%
  \BibitemOpen
  \bibfield  {author} {\bibinfo {author} {\bibfnamefont {M.}~\bibnamefont {Van
  Den~Broeke}}, \bibinfo {author} {\bibfnamefont {J.}~\bibnamefont {Bamber}},
  \bibinfo {author} {\bibfnamefont {J.}~\bibnamefont {Ettema}}, \bibinfo
  {author} {\bibfnamefont {E.}~\bibnamefont {Rignot}}, \bibinfo {author}
  {\bibfnamefont {E.}~\bibnamefont {Schrama}}, \bibinfo {author} {\bibfnamefont
  {W.~J.}\ \bibnamefont {Van De~Berg}}, \bibinfo {author} {\bibfnamefont
  {E.}~\bibnamefont {Van~Meijgaard}}, \bibinfo {author} {\bibfnamefont
  {I.}~\bibnamefont {Velicogna}},\ and\ \bibinfo {author} {\bibfnamefont
  {B.}~\bibnamefont {Wouters}},\ }\bibfield  {title} {\bibinfo {title}
  {Partitioning {Recent} {Greenland} {Mass} {Loss}},\ }\href
  {https://doi.org/10.1126/science.1178176} {\bibfield  {journal} {\bibinfo
  {journal} {Science}\ }\textbf {\bibinfo {volume} {326}},\ \bibinfo {pages}
  {984} (\bibinfo {year} {2009})}\BibitemShut {NoStop}%
\bibitem [{\citenamefont {Kypke}\ \emph
  {et~al.}(2025{\natexlab{b}})\citenamefont {Kypke}, \citenamefont {Ashwin},\
  and\ \citenamefont {Ditlevsen}}]{model_code}%
  \BibitemOpen
  \bibfield  {author} {\bibinfo {author} {\bibfnamefont {K.}~\bibnamefont
  {Kypke}}, \bibinfo {author} {\bibfnamefont {P.}~\bibnamefont {Ashwin}},\ and\
  \bibinfo {author} {\bibfnamefont {P.}~\bibnamefont {Ditlevsen}},\ }\href
  {https://doi.org/10.5281/zenodo.17288831} {\bibinfo {title} {Model code and
  figure generating scripts for ``{C}haotic variability in a model of coupled
  ice streams"}},\ \bibinfo {howpublished} {{Z}enodo,
  https://doi.org/10.5281/zenodo.17288831} (\bibinfo {year}
  {2025}{\natexlab{b}})\BibitemShut {NoStop}%
\end{thebibliography}%

\end{document}